
%
%
\def\minstep#1{\ifcase#1 900\or 830\or 750\or 500\fi\relax}

\font\twrm=cmr12
\font\ninerm=cmr9
\font\twi=cmmi12
\font\ninei=cmmi9
\font\twsy=cmsy10 scaled \magstep2
\font\ninesy=cmsy9
\font\sevensy=cmsy7
%

\font\twbf=cmbx12
\font\ninebf=cmbx9
\font\twsl=cmsl12
\font\twex=cmex10 scaled \magstep2
\font\twit=cmti12
\font\nineit=cmti9
\font\twtt=cmtt12
\textfont0=\twrm
\scriptfont0=\ninerm
\scriptscriptfont0=\sevenrm
\def\rm{\fam0\twrm}
\textfont1=\twi
\scriptfont1=\ninei
\scriptscriptfont1=\seveni

\textfont2=\twsy \scriptfont2=\ninesy \scriptscriptfont2=\sevensy
\def\cal{\fam2\twsy }
\textfont3=\twex \scriptfont3=\twex \scriptscriptfont3=\twex
\textfont4=\teni \scriptfont4=\teni \scriptscriptfont4=\teni

\newfam\itfam  \textfont\itfam=\twit
\newfam\slfam  \textfont\slfam=\twsl
\newfam\bffam  \textfont\bffam=\twbf
                                       \scriptfont\bffam=\ninebf
                                       \scriptscriptfont\bffam=\sevenbf
\newfam\ttfam  \textfont\ttfam=\twtt
%
%
\topskip=12pt
\newdimen\headersize \headersize=40pt
\baselineskip=15.0pt plus .1pt \parskip=5pt plus .3pt
\hsize=6.5in \vsize=8.88in \advance\vsize by -\headersize
\hoffset=0in \voffset=0.0in
\newif\iffirstpage
\let\nis=\nointerlineskip
\def\makeheadline{\ifnum\pageno=1\else\nis\setbox0=\pagehead\dp0=0pt\box0\nis\fi}
\def\makefootline{\ifnum\pageno=1\nis\setbox0=\pagefoot\dp0=0pt\box0\nis\fi}
\let\plainadvancepageno=\advancepageno
\def\advancepageno{\plainadvancepageno\global\firstpagefalse}

\def\pagehead
  {\vbox to \headersize {%
   \setbox0=\line {\ninebf\hfil \folio}%
   \ht0=8.5pt \dp0=3pt \box0\vfil}}

\def\pagefoot
  {\vbox to \headersize {%
   \setbox0=\line {\hfil\nineit\folio}%
   \ht0=8.5pt \dp0=0pt \vfil \box0}}
%
%

\def\item{\vskip 4pt\noindent}
\def\smitem{\vskip -3pt\noindent}
%
%

\def\D{\Delta}

\def\refform#1#2#3#4{{\twrm #1,} {\twsl #2,\/} {\twbf #3,} {\twrm #4.}}
\def\refformb#1#2#3#4{{\twrm #1,} {\twsl #2,\/} {\twbf #3,} {\twrm #4}}

\def\refbook#1#2#3#4{{\twrm #1}{\twsl #2,\/} {\twrm #3,} {\twrm (#4).}}

\twrm
%

%

%
%
\def\abstrPRE{%
We study how doping destroys the AF order in the layered cuprates
within the framework of the charge--transfer insulator concept.
We use the criterion of stability of the AF background
to show  that the stability problem is
one of the main issues in any correspondence between results for the
$t-J$ model and, say, the three--band model for
the lightly--doped layered oxides.  Provided a
phenomenological conduction
band is chosen to satisfy the criterion of stability,
a detailed picture of how dopants influence the spin
wave spectrum at $T=0$ is presented. The critical
concentration $x_c$ for the destruction of the AF
long range order is due
to the Cherenkov effect when the Fermi velocity first exceeds
the spin wave velocity. We then discuss the overall
spectrum of spin excitations
and find that the spin wave attenuation for $x < x_c$, $T=0$
due to Landau damping appears in the range of
magnon momenta $k(x) = 2 m^* s \pm \alpha \sqrt{x}$.
We also argue that in the
presence of  superconductivity, the Cherenkov effect is
eliminated due to the gap in the spectrum. This may restore the role
of the AF fluctuations as the main source of dissipation at the
lowest temperatures.}
%
\baselineskip = 24pt
\twrm

\centerline{\twbf Evolution of Magnetic Properties of Lightly
Doped Copper Oxides}

\tenit\centerline{\twrm L.~P.~Gor'kov $\rm ^{(a)}$,
V. Nikos Nicopoulos $\rm ^{(b)}$ and Pradeep Kumar $\rm ^{(b)}$}
\centerline{$\rm ^{(a)}$National High Magnetic Field Laboratory,
Florida State University, 1800 E. Paul Dirac Dr.,
Tallahassee, FL 32306}
\centerline{and L.~D.~Landau Institute for Theoretical Physics,
the Academy of Sciences of Russia,}
\centerline{2 Kosygina Street, 11733V Moscow,
Russia}
\centerline%
{$\rm ^{(b)}$Physics Department, University of Florida,
Gainesville, FL 32611}

\vskip 1cm
\centerline{\twbf ABSTRACT}

\twrm
\abstrPRE
\vskip 1cm\noindent%
PACS \#: 74.70Vy, 75.30.-m, 73.20.Dx
\vfill\break
\twrm
We address the issue of how the conduction and
magnetic properties of the HTSC oxides evolve with doping starting
from Mott's insulator state.  The system best studied
experimentally so far is La$_{2-x}$Sr$_x$CuO$_4$. The
properties of the AF
insulating state for the parent compound, La$_2$CuO$_4$,
are by now reasonably
understood in terms of the 2D Heisenberg model (see [1]
for a
review).  The origin of magnetism is commonly ascribed to the copper
$d^9$ configuration (i.e. one hole in the closed $d^{10}$
Cu shell).  The
Hund's rule in the atomic limit would provide $S = 1$ for the
configuration $d^8$, which is costly in energy (7--10 eV), so that the
so-called ``Hubbard $U$", the repulsion between two holes located at
the ``same Cu site", can safely be taken equal to infinity
($U_{Cu} =\infty$ ).

An enormous literature exists by now regarding properties of
lightly doped 214--materials (La$_{2-x}$Sr$_x$CuO$_4$).
The vast majority of
theoretical papers treat the hole motion in terms of the $t-J$ model.
In this model doping is represented by the {\twit removal} of
charges originally
located on Hubbard centers.  On the contrary, as is known, the
actual situation is the opposite
(at least for 214 compounds) -- Sr$^{2+}$ ions
{\twit add} charge (holes) into
the system.  The resolution of the seeming contradiction with the
assumption of infinite U comes if one remembers that the  new
materials
are charge-transfer insulators [2], i.e., in addition to Hund's
energy, there exists a lower energy scale.  Indeed, transfer of
electrons from an oxygen site, $\epsilon_p$, to close up the
copper $d$-shell
($d^{10}$) $\epsilon_d$, costs in energy only
$\left|\epsilon_p-\epsilon_d\right| = 2\ {\rm to}\ 2.5$ eV
(see [2] for a review).
Without doubt, in the complicated  La$_2$CuO$_4$ unit
cell  all orbitals become mixed, or hybridized.
Although the description in terms of ionic copper or oxygen orbitals
serves mostly as a convenient language, it contains the physics
underlying the essential properties of these materials.

There is a belief that the above-mentioned controversy concerning
adding holes by doping in real materials and its simulation in
terms of ``holes" formed in the half-filled background of the $t-J$
model may be eliminated by a one-to-one mapping
in terms of the Zhang-Rice singlet picture [3].
In this view
a hole introduced by Sr$^{2+}$ goes to oxygen orbitals
where it forms
a singlet bound state (in the three-band model [4]) of the oxygen
orbitals with one of the neighboring localized Cu spins.  Such a
picture is  not obvious.  Its applicability has already been the
subject of a discussion [5,6].  Our considerations
have also led us to the conclusion that any search
for such a correspondence simply misses the main physics here.

A more crucial question
is whether the dynamic properties
of mobile holes in HTSC materials and their impact on the magnetic
behavior can nevertheless be simulated by the $t-J$ model.  It seems
obvious
that the physics becomes richer if the extra
electronic degrees of freedom connected with the lower oxygen
levels are taken into account.  On the other hand, at low doping
and low temperatures some characteristics of doped oxides may be
model independent and can be treated 
phenomenologically.
Therefore, in the first part of this paper we discuss these
general features as they can be derived for the lightly doped
Mott state from the charge--transfer point of view.
For this purpose, there is no need to go through all the
complications of, say, the three band model [4, 6, 7].
Instead, we adopt a more phenomenological point of
view.  Namely, we assume that, in addition to the localized Cu
states where the electron correlations are most crucial, the
manifold of other (``oxygen") orbitals provides a set of itinerant
bands of a common character to which the doped holes go.  In such
an approach [7]  the standard Anderson
Hamiltonian is adequate for the description of the behavior of
holes:
$$%
H = \sum_{i\sigma} \epsilon_d d_{i\sigma}^{\dagger} d_{i\sigma}
+ \sum_{\vec p\sigma} \epsilon(\vec p)
             a_{\vec p\sigma}^{\dagger} a_{\vec p\sigma}
+ t_0 \sum_{i\vec p} \left(d_{i\sigma}^{\dagger} a_{\vec p\sigma}
e^{i\vec p\cdot\vec r_i}
+{\rm h.c.}\right)\eqno{(1)}
$$

Here $\epsilon_d$ is the localized (Cu) level, $\epsilon(\vec p)$
the dispersion law of
the itinerant band.  Let us also assume that
$\epsilon(\vec p) \sim p^2/2m$ near $\vec p=0$.
Then $W = 4/(m a^2)$ is a scale for the bandwidth
(we use units such that $\hbar = 1$ and $a$ is the lattice
constant),
while $t_0$ is responsible for hybridization,
i.e., for
transitions between localized and delocalized states separated by
the charge transfer gap $\epsilon_p - \epsilon_d \sim \Delta$,
and double occupancy of the copper sites is forbidden
($U_{Cu} = \infty$).
The virtual processes then immediately
provide a ``Kondo--like" exchange interaction between the spin of a
conduction hole and the spins of localized levels $\vec S_i$:
$$%
j'(\vec S_i\cdot\vec\sigma)\quad\quad ;%
\quad\quad j'= {2 t_0^2\over\Delta} \eqno(2)
$$

In higher order the Heisenberg exchange between
nearest neighbor localized
spins appears:
$$%
H_{\rm ex} = \sum_{\langle i,k\rangle} J_{ik} \vec S_i\cdot
\vec S_k \eqno(3)
$$
The nearest neighbor exchange interaction is
nonvanishing for this model
only for a finite bandwidth $W$. In the case that $W\ll\Delta$
we have
$$
J_{ik} = J =  { t_0^4\over 2\Delta^3} \left({W\over\Delta}\right)^2
\eqno(3')%
$$


With both $\Delta$  and J known experimentally
($\Delta\simeq 2.0$ eV, $J\simeq 0.15$ eV)
one estimates (assuming $W\simeq\D$)
for $\left(t_0^2/\Delta^2\right)\simeq 0.4$.
(The
three-band model may involve different factors in eqs. (1,2);
see refs. [4,5] and below).

The value for $J\simeq   1500$ K is rather large and the
common belief is
that to describe doping, at least for small enough $x$, the
antiferromagnetic background is always to be taken into account
first.  Rewriting the effective Hamiltonian (2) and (3) in the
momentum space:
$$%
H_{\rm eff} = \sum_{\vec p\sigma} \epsilon(\vec p)
             a_{\vec p\sigma}^{\dagger} a_{\vec p\sigma}
+ {j'\over \sqrt{N}} \sum\vec S_{\vec q}%
\cdot\left(a^\dagger_{\vec p -\vec q\rho}
\vec\sigma_{\rho\rho'} a_{\vec p, \rho'}\right)
+\sum J(\vec q) \vec S_{\vec q}\cdot\vec S_{-\vec q}\eqno(4a)
$$
and after introducing (at $T =0$ in the $2d$ case) the average
staggered magnetization:
$\vec\mu = \mu_B\langle\vec S_{\vec Q_0}\rangle
\neq 0$, where $\vec Q_0 \equiv (\pi/a,\pi/a)$, the total Hamiltonian
(4a) may then be
reduced to a simpler form, since  for small concentrations $x$  only
the bottom
of the conduction band is important.   In the vicinity of $\vec p
= 0$
and in the presence of a staggered magnetization   the behavior of
a single
hole is described by the Hamiltonian:
$$%
H_{\rm eff} = %
\sum_{\vec p\sigma}{p^2\over 2m} a_{\vec p\sigma}^{\dagger} %
                            a_{\vec p\sigma} %
+ {1\over \sqrt{N}} \sum_{\vec p,\vec q}%
                         \left[ j'\ \vec S_{\vec q}
                         + i B \vec f(\vec p,\vec q)\cdot \vec q\
\ \vec j_s\right]\cdot%
                       \left(a^\dagger_{\vec p-\vec q\rho} %
\vec\sigma_{\rho\rho'} a_{\vec p , \rho'}\right)
\eqno(4b)
$$

The coefficient  B is $a^2j'^2/8 W$ (we assumed
$W\gg j', J$ for simplicity),
and the momentum dependent form factor
$%
\vec f \equiv \vec p - {\vec q\over 2}
$.
The term  with
$%
\vec j_s \equiv (\vec \mu\times \vec S_{\vec Q_0-\vec q})/\mu_B
$
appears as a perturbation
due to opening of an AF gap in the new band spectrum at
$\vec p = \vec Q_0/2$.  One sees
that eqn. (4b) has the same form as that derived
in [8,9] from
the $t-J$ model in the
nonlinear $\sigma$ model approximation.
The distinction between
the two pictures is that in the above the number of localized spins is
conserved.  Secondly, in such an approach there are no special
relation between $m$ (or $W$), $J$ and
$j'$  and, finally, doping adds new
fermionic degrees of freedom, which are now responsible for
conduction.

At the derivation of eq. (4b) (near $\vec p = 0$) it was assumed
that only
the long wave components of spin degrees of freedom are relevant.
Those are the spin waves and can be introduced either
phenomenologically, or derived explicitly by use of the Dyson -- %
Maleev transformation [10].  For our immediate needs we write down
only the second term in eq. (4b):
$$%
{j'\over \sqrt{N}}\ %
\sum_{\vec p,\vec q}  {\sqrt{q a}\over 2^{3/4}} %
\left[%
\left(\alpha_{\vec q} + \beta^\dagger_{-\vec q}\right)
a^\dagger_{\vec  p - \vec q\downarrow} a_{\vec p\uparrow}
+ \left(\alpha_{\vec q}^\dagger + \beta_{-\vec q}\right)
a^\dagger_{\vec  p - \vec q\uparrow} a_{\vec p\downarrow}%
\right].%
\eqno(5)
$$

Together with the Hamiltonian for spin-waves
$$
H' = \sum \Omega_{\vec q} \left(\alpha^\dagger_{\vec q}
                                \alpha_{\vec q}
+\beta^\dagger_{\vec q} \beta_{\vec q}\right)
\eqno(6)
$$
one arrives at the well-studied problem of an electron
interacting with
long wave excitations having the linear dispersion law
$\Omega(\vec q) = s q$ (see [11] for a review).
This
allows one  to reexamine the self-consistency of the basic assumption
that the motion of holes takes place in the AF background.  The
problem is that the hole described by Hamiltonian (4b) and
interacting with spin waves may be unstable with respect to a sort
of polaron formation.  If spin frequencies remain low compared with
the hole bandwidth ($W>>J$) the problem may be investigated
quasi--classically
in  the spin wave occupation numbers ($\alpha^\dagger\alpha,
\beta^\dagger\beta\sim j'^2/(J^2qa) >> 1$).  It is possible
to minimize the sum of eqs. (4b) and (6) to eliminate
the spin wave operators. The result is the familiar Deigen-Pekar
functional
for the ground state of the hole ``dressed" by spin fluctuations:
$$%
H = {1\over 2m}\left\{\int\sum_{\sigma}\left|\nabla \psi_\sigma\right|^2
- - -4 c \left|\psi_{\downarrow}\right|^2\left|\psi_{\uparrow}\right|^2
d^2\vec r\right\}
\eqno(7)
$$
where $\psi$  is the hole wave function and
$$%
c \equiv {m j'^2 (q a) a^2\over 2\sqrt{2}\ \Omega_q},
\eqno(8)
$$
where $\Omega_q$ is the spin
wave energy.

In ordinary semiconductors the functional (7)
determines the height of the barrier for formation of self-trapped
states.  The property specific to the $2d$ physics is
that in this case such
a barrier does not exist.  Instead, as has first been shown in [12]
 the system either remains stable if
$$%
c < 2.88
\eqno(9)
$$
or has a saddle point (i.e. the functional (7) has no minimum) if
$c$
exceeds the limit (9).  The third term in (4b) does not add much
to these considerations.  Provided the parameters of the model are
such that AF background remains essentially intact,
this term may provide
a helical structure in the doped antiferromagnet
[9].

Having this criterion, we first discuss its
implications for, say, the three--band model.  We have
considered whether this model can be reduced to the form of eq.
(4a).  (The results will be published in detail
elsewhere).  It turns out that the truncation of the three-band
Hamiltonian done in refs. [3] and [5] to a sole
singlet-like component
essentially oversimplifies the results.  Actually, this model,
after the nonzero staggered magnetization is taken into account,
becomes highly degenerate and possesses dispersionless bands.  To
eliminate these artifacts, it is necessary to account for
a number of realistic features like
the interaction between two holes placed on an oxygen
site (we assume a finite Hubbard $U_O$, for double occupancy of
oxygen sites).  Then we would obtain:
$$%
j' = -\ {4t_0^2 U_O\over\Delta(\Delta+ U_O)} < 0
\eqno(10)%
$$
and, for the effective mass of the lowest band with dispersion
$$%
\left(m^*\right)^{-1} = {2 a^2\left(1-(\mu/\mu_B)^2\right)t}\ \ ,
\eqno(11)%
$$
where $t = t_0^2/2\Delta$.

For the three band model with an ``on--oxygen" $U_O$,
the antiferromagnetic
interaction (3) turns out to be
$$%
J = 4\ {t_0^4\over\Delta^2(2\Delta+U_O)}
\eqno(3^{\prime\prime})%
$$

The criterion for stability for this band would then be
$$%
{1\over\left(1-(\mu/\mu_B)^2\right)}\ \ {U_O^2\over t_0^2}\ \ %
{\Delta(2\D + U_O)\over (\D+U_O)^2} < 2.88
\eqno(12)%
$$.

We see that the criterion depends on all three parameters
$t_0, \Delta$ and
$U_O$, so that (as before) in eqs. (3$^{\prime\prime}$, 9) we need
an additional experimental parameter
to determine whether it is satisfied or not.
If, however, one starts from the model assumed in
[7] ($W\gg\D$),
the band width, $W \gg j'\gg J$, then $c\ll 1$ and   the
criterion (9) for the stability of the AF state is fulfilled.
We may analyze the $t-J$ model results from the same point of
view. While the phenomenological Hamiltonian (4b) has the same
form in both cases as it is noted above, the parameters are
quite different. For the orthodox $t-J$ model $t$
is expected to be
large $t\gg J$. To get the kinetic energy, one is to account
at least partially for fluctuations. The issue of the bandwidth
remains a controversial one (see ref. [9]), however it seems
that $W$ is of the order of $J$.  Though no adiabatic approximation
is justified any more, the criterion would still point toward
instability
if $W\sim J\ll j'\sim t$.
Therefore, in our opinion, the 
problem is not at all
whether the one-to-one correspondence between the three--band picture
and the $t-J$ model can be established. It is that
if $j'\gg J$ the  system may become
unstable and, as it is easy
to estimate, the instability would be resolved by the formation of
large magnetic ``polarons" (or textures) of the size $r_0$, such that
$r_0^2/a^2\sim j'/J \gg 1$, so that any
correspondence would lose any meaning.  Indeed, the
criterion (9) can now be understood better from this end, at
least qualitatively.  Let us overlook for a moment the
kinetic energy term in eqn. (4b).
Then the interaction $j'$ tends to align
the surrounding spins in a parallel way.  It produces a ``potential
well"
of depth of the order of $j'$ to a distance less than $r_0$.
The  well has
a finite size $r_0$ because there is an energy cost of the order of
$J(r_0/a)^2$
due to the destroyed  AF state within the well.
One now realizes that the
criterion (9) indicates the provision that due to the
kinetic energy terms one may
or may not have a bound state in this effective potential well.
The onset of an instability may mean either the creation of a
number of textures or even ``phase separation" [13].

Assume now that, in accordance with eqn. (9),
with the appropriate values for the parameters  $W, j', J$
the AF ground state
remains stable. We address now another problem, namely
how doping would  destroy the AF order ($T\equiv  0$).  For finite
concentration of dopants sitting at the bottom of the band
(near $\vec p=0$) there are corrections
to the spin
wave spectrum shown diagrammatically in Fig. 1.
The straightforward calculation of the polarization operator
gives rise to the following expression for the renormalized
spin wave velocity $\bar s$ ($\vec q\rightarrow 0$):
$$%
\bar s = s + {j'^2 a\over  \sqrt{2}\pi W}
\int {(\vec q\cdot\vec v_F) \ d\phi
\over q s - (\vec q\cdot\vec v_F) - i \delta}
\eqno(13)
$$

(Recall that the long range AF order implies (and vice versa) a
Goldstone mode in $\vec q$ at $q\rightarrow 0$).
The expression in brackets can be calculated explicitly.
For $v_F < s$ one has:
$$%
\bar s = s\ \left\{1 + {c\over 2\pi}
\left[{1\over\sqrt{1 - \left(v_F/s\right)^2}} - 1%
\right]\right\},
\eqno(13')
$$
where $c$ is as  in eqn. (8).
The remarkable feature in eqn ($13'$) is that the attenuation of
the spin waves at $q\rightarrow 0$ takes place abruptly in the form
of the inverse square root singularity at $v_F \ge s$.
The mechanism (the Cherenkov effect) is so pronounced due to the
dimensionality ($d = 2$).
We can now write down the critical concentration $x_c$ of
dopants destroying the AF long range order by this mechanism
(the doping parameter  $x = (a p_F)^2/2\pi$):
$$%
x_c = {(a m^* s)^2\over 2\pi} \sim \left({J\over W}\right)^2\ll 1
\eqno(14)%
$$
A few comments are in order in connection with
eqs. (13), ($13'$) and (14).
First, one sees from (13), ($13'$) that at
very low concentrations $v_F < s$ the stiffness of the spin
wave mode at very low $q$ increases with doping.
This is unexpected, but it seems it can be interpreted as a
reduction of quantum fluctuations due to doping for a
stable AF background.
Let us now discuss in more detail how doping would influence the
AF state as a whole. The short wavelength part of the spin wave spectrum
remains
of course untouched by
either finite temperature ($T < \omega(\vec q), T\ll J$),
or doping (at $x\ll 1$). At $T = 0$ the role of doping
is then clarified by a more careful study of the divergence in Fig. 1
at arbitrary $\vec q$. Here we summarize the results.
At $x\rightarrow 0$, the attenuation in $\omega(\vec q)$ first
appears around $q^* \equiv 2 m s$. At finite $x<x_c$ the imaginary
part in $\omega(\vec q)$ is non--zero in the range
$%
2 m s - 2 p_F < q < 2 m s + 2 p_F
$ (Landau damping).
Taking $\left(\xi^*\right)^{-1} = 2 \pi q^* = 4\pi (m s + p_F)$,
the beginning of attenuation from the side of larger $q$,
we get what we might call a coherence length
$$%
\left(\xi^*\right)^{-1} = 4\pi\left( m s +
 \sqrt{2\pi}\ {x^{1/2}\over a}\right)
\eqno(15)%
$$
Comparing this with $2 \pi q_T\sim 2\pi T/s$ we see that the
Heisenberg like AF fluctuations would become
``saturated" at
$$%
T^* \simeq 2\sqrt{2\pi}\  {s x^{1/2}\over a}
\eqno(16)%
$$

In the presence of disorder introduced by dopants (Sr$^{+2}$--ions)
the conduction holes experience scattering by defects. Taking the
mean free path $l$ for holes into account
one would get an additional source of attenuation of the spin waves.
At $q l \ll 1$ this effect on the diagram in Fig. 1 is known to
be to substitute for the factor under the integral
in eq. (13):
$$%
\int {(\vec q\cdot\vec v_F) \ d\phi
\over q s - (\vec q\cdot\vec v_F) - i \delta} \rightarrow
\int \ d\phi\ {-i{\cal D} q^2\over \omega + i{\cal D} q^2}
\eqno(17)%
$$
For $\omega > {\cal D} q^2$ we get
$$%
\omega(\vec q) = s \left(q - i {c\over 2\pi} {{\cal D} q^2\over s}%
\right)
\eqno(18)%
$$
as it should be in the hydrodynamic regime. This result ignores
any spin--flip scattering by defects.

Both the Green's functions
({\twit i.e.\/} $m$) and the vertices
({\twit i.e.\/} $j'$) in the diagram shown in Fig.1 are renormalized
by quantum corrections in accordance with the Hamiltonian (5).
It is easy to verify that the main corrections are of the order
of $c$, {\twit i.e.\/} they would not seriously change the estimate
(16) provided (9) is satisfied.

To summarize, the spin density wave spectrum which is responsible
for the spin fluctuations in doped systems acquires an attenuation at
$q^* = 2 m s $ (Landau damping).
There is a sharp threshold due to the Cherenkov effect
(eqs. (13--$13'$)), which describes the disappearance of the AF
long range order at $T=0$ as manifested by the
dissappearanceof the Goldstone
mode at $q\rightarrow 0$. Most remarkable is that for the
Cherenkov effect to show up, a sharp Fermi surface at zero
temperature is needed. Therefore if the superconducting gap
were present for $x> x_c$, the mechanism ($13'$) would disappear
as if a restoration of some AF order had taken place.
Of course, the effects of the mean free path
for hole scattering from defects would also smear out the Fermi
surface edge. Unfortunately, the whole analysis carried out for $T=0$
cannot be immediately extended to an analysis of the system's
temperature behavior, and the points above need further investigation.

One of us (LPG) acknowledges the extensive discussions with
J.~R.~Schrieffer at the beginning of this work and the
valuable comments by E.~I.~Rashba on the stability problem
in semiconductors. LPG was supported by the
National High Magnetic Field Laboratory through the NSF
cooperative agreement \# DMR--9016241 and the State of Florida.
VNN and PK were supported by a grant from U. S. D. O. E.,
DEF G05--91--ER45462.

\vfill\break
\centerline{\twbf REFERENCES}
\item%
[1]  \refbook{S. Chakravarty in }
{High $T_c$ Superconductivity, K.~S.~Bedell et al., eds.}
{p. 132}{Addison-Wesley, 1990}
\smitem
[2]  \refbook{T. Timusk and D.~B.~Tanner in }{Physical Properties of High
$T_c$ Superconductors, vol. I, D.~M.~Ginsberg, ed.}
{p. 339}{World Scientific, 1989}
\smitem
[3]  \refform{F.~C.~Zhang and T.~M.~Rice}{Phys. Rev. B}{37}%
{3759 (1988)}
\smitem
[4]  \refform{V.~J.~Emery}{Phys. Rev. Lett.\/}{58}{2794  (1987)}
\smitem
[5]  \refform{V.~J.~Emery and G.~Reiter}{Phys. Rev. B\/}{38}{11938 (1988)}
\smitem
[6]  \refformb{F.~C.~Zhang and T.~M.~Rice}{Phys. Rev. B\/}{41}{7243 (1990)};
\refform{V.~J.~Emery and G.~Reiter}{Phys. Rev. B\/}{41}{7247 (1990)}
\smitem%
[7]  \refformb{L.~P.~Gor'kov and A.~V.~Sokol}{Sov. Phys. JETP\/}
{46}{420, (1987)};
\refform{L.~P.~Gor'kov and A.~V.~Sokol}{Journ. Physique}{50}
{2823 (1989)}
\smitem%
[8] \refform{Boris I.~Shraiman and Eric D.~Siggia}{Phys. Rev. Lett.\/}
{61}
{467 (1988)}
\smitem%
[9]  \refform{Boris I.~Shraiman and Eric D.~Siggia}{Phys. Rev. B\/}
{42}
{2485 (1990)}
\smitem%
[10] \refformb{F. J. Dyson}{Phys. Rev.\/}{102}{1217 (1956)};
\refform{S.~V.~Maleev}{Sov. Phys. JETP\/}{6}{776 (1958)}
\smitem%
[11]  \refbook{E.~I.~Rashba in }{Excitons, E.~I.~Rashba and
M.~D.~Sturge, eds.\/}{p. 543}{North--Holland, 1982}
\smitem%
[12]  \refformb{E.~I.~Rashba}{Sov. J. Low Temp. Phys.}{3}{254 (1977)};
\refform{R. Y. Chiao, E. Garmire and C. H. Townes}
{Phys. Rev. Lett.}{13}{479 (1964)}
\smitem%
[13]  \refform{V.~A.~Kashin and E.~L.~Nagaev}{Sov. Phys. JETP}{39}{1036 (1974)}
\smitem%
\vfill\eject
\centerline{FIGURE CAPTION}

{\twbf Fig. 1:}  Diagram that contributes to the renormalization
of the spin wave velocity to lowest order.

\bye